# Not All Scaffolds Are Equal: How Initiation Mode Determines EMME Effectiveness in Debugging

Anahita Golrang [a], Kshitij Sharma [a], Halszka Jarodzka [b], Senne Van Hoecke [b]

[a]Department of Computer Science, Norwegian University of Science and Technology, 7034 Trondheim, Norway
[b]Expertise Centre for Education, Open University of the Netherlands, the Netherlands

## Abstract

**Background:** Adaptive learning technologies increasingly rely on real-time physiological analytics to trigger instructional support automatically, yet how system-driven decisions interact with learners' ongoing problem-solving processes remains poorly understood. Eye Movement Modeling Examples (EMME) have shown promise as attention-guidance tools, but have been studied predominantly as static instructional materials rather than as adaptive scaffolds whose timing and initiation control can vary. Critically, the question of whether EMME support should be initiated by the teacher, the learner, or an automated system has received no direct empirical attention, despite having fundamental implications for scaffold alignment and learner agency.

**Objectives:** This study investigates whether scaffold initiation mode shapes EMME effectiveness in novice programmers' debugging, and specifically whether automated triggering based on a single physiological indicator of low mental effort is a viable basis for adaptive scaffold delivery.

**Methods:** A between-subjects experiment was conducted with 120 undergraduate computer science students randomly assigned to one of four conditions: teacher-initiated, learner-initiated, automated, or no-scaffold control. Participants completed ten python debugging tasks while eye-tracking data, video interaction logs, and performance scores were recorded.

**Results and Conclusions:** All EMME conditions outperformed the control. However, human-mediated initiation, whether teacher or learner, consistently produced higher performance than automated triggering and more integrative engagement with the EMME material. Automated triggering, based on sustained low pupillary activity, was associated with disruptive behavioral patterns suggesting mistimed delivery. EMME also eliminated the performance advantage of prior programming knowledge across all initiation modes. These findings establish scaffold initiation timing and control as critical and previously unexamined design variables for EMME and adaptive learning technologies more broadly, and demonstrate that a single low-effort physiological threshold is insufficient as a trigger criterion for complex problem-solving support.

**Practitioner notes**

**What is already known about this topic**

- Debugging is a cognitively demanding, ill-structured task in which novice programmers frequently struggle to allocate attention strategically and seek help at appropriate moments.
- EMME make expert visual attention and problem-solving strategies observable to learners and have shown generally positive effects on attention guidance and learning outcomes across domains.
- EMME findings in complex tasks such as programming debugging remain mixed, with inconsistencies attributed to task complexity, learner expertise, and instructional design factors, but the role of scaffold initiation timing and control has not been examined.
- Adaptive learning systems increasingly use real-time physiological data to trigger support automatically, but the conditions under which such triggering supports versus disrupts ongoing learning are not well understood.

**What this paper adds**

- EMME functions not only as a modelling tool but as an adaptive regulatory scaffold that supports learners' monitoring and control during complex problem solving.
- Who initiates scaffolding – teacher, learner, or system – is a critical design variable, with human-mediated initiation consistently outperforming automated triggering on debugging performance.
- Productive regulatory engagement, reflected in strategic task-video transitions, distinguishes effective scaffolding from automated triggering, which induced disruptive behaviours such as excessive pausing and backward navigation.
- EMME eliminated the performance advantage associated with prior programming knowledge, suggesting its potential to promote more equitable outcomes across varying skill levels.

**Implications for practice and/or policy**

- Both teacher-identified and learner-requested support are viable and equivalently effective delivery modes for EMME, giving instructional designers flexibility in implementation without sacrificing effectiveness.
- The equalizing effect of EMME across expertise levels makes it a promising tool for supporting equity in introductory programming education.
- Designers of adaptive learning systems should not rely on a single physiological metric of low mental effort as a trigger for complex problem-solving support: low effort may reflect productive fluency rather than disengagement, and triggering support in such moments risks interrupting rather than facilitating learning.
- Future adaptive systems should combine behavioral, physiological, and task-based indicators to build richer, more context-sensitive triggering algorithms before deploying automated scaffolding in complex domains.

# 1 | INTRODUCTION

## 1.1 | The challenge of automated support timing in adaptive learning systems

Adaptive educational technologies increasingly rely on real-time analytics to determine when instructional support should be delivered. Advances in eye-tracking, physiological sensing, and learning analytics have made it technically feasible to trigger scaffolding dynamically during learners' activity. However, the growing capacity for automated intervention has outpaced empirical understanding of how such system-driven decisions interact with learners' self-regulatory processes during complex problem solving [1, 2]. This creates a fundamental design tension: automated systems can respond faster than teachers or learners, but they may respond at the wrong moment, to the wrong signal, or in ways that interrupt productive engagement rather than facilitating it.

## 1.2 | Debugging as a critical domain for studying scaffold initiation

Debugging is an ill-structured task that places heavy demands on learners' regulation, making it an ideal context to examine how different scaffold-initiation modes affect learning. Ill-structured tasks, such as programming debugging, lack clear solution paths and require learners to coordinate domain knowledge with sustained regulation of strategies and effort [3]. Debugging is an ill-structured task since it requires learners to evaluate program behavior, generate and test hypotheses, detect impasses, and decide whether to persist, revise strategies, or seek support. Research consistently shows that novice programmers struggle to isolate errors, prioritize debugging strategies, and transfer their skills across contexts [4]. As a result, debugging provides a theoretically and practically relevant context for examining how different scaffold initiation modes interact with learners' regulation.

To address these challenges, computer-assisted learning tools such as program visualization interfaces [5] and linked representations [6] have been developed to externalize code behavior. While these tools help externalize program behavior, they often fall short of modeling expert problem-solving processes. Novices may see what happens in the code, but not how experts allocate attention, diagnose problems, or decide on next steps. This limitation highlights the need for scaffolds that externalize not only outcomes, but also expert perceptual and strategic processes.

## 1.3 | EMME as attention guidance and the limits of prior work

Although Eye Movement Modeling Examples (EMME) support attention and strategy modeling [7], mixed findings [8, 9] suggest that their effectiveness depends on how and when they are introduced. EMME are instructional videos that overlay an expert's gaze patterns onto a screen recording of their task performance [10]. Grounded in observational learning theory [11], EMME direct learners' attention to task-relevant elements and model effective strategies [8, 12]. Their benefits are further explained through cognitive load theory, which predicts that expert gaze cues reduce extraneous load during task performance [12, 13, 14]. Empirical work across domains, including medical diagnostics and multimedia learning, demonstrates that EMME can accelerate attention to critical areas and improve performance [15, 16, 8, 17].

Despite these promising results, the EMME literature shows inconsistent findings [18, 13, 19], particularly for complex cognitive tasks such as programming debugging [17, 8]. Recent systematic reviews help clarify when EMME tends to succeed or fail. Tunga and Cagiltay [9] found that effectiveness is moderated by task characteristics: EMME produces stronger results when tasks have a strong visual component, low distraction, and a non-procedural nature. Paradoxically, EMME has been applied more frequently to cognitive than perceptual tasks, and this mismatch likely contributes to the field's inconsistent findings. Programming debugging exemplifies this challenging profile; it is procedural, involves navigating dense and potentially distracting source code, and demands sustained hypothesis-driven reasoning [17]. These task characteristics alone would predict mixed results, yet they do not fully account for observed variability, as some EMME interventions in cognitively demanding domains have still produced meaningful learning gains [8].

We argue that a systematic but previously overlooked source of this variability is scaffold initiation mode. Most prior EMME studies have delivered the intervention at fixed, predetermined points either at the start of a task, be-tween tasks, or as a continuous overlay, without examining how the timing and control of delivery interact with learners' ongoing engagement. When EMME is deployed as a feedback tool rather than as a static instructional material, control over its initiation becomes a central design dimension. Inconsistencies in the EMME literature may therefore reflect not a limitation of gaze-based modeling itself, but a failure to account for when and by whom the scaffold is activated. The present study addresses this gap by systematically manipulating initiation mode as the primary independent variable, guided by two core questions:

1. **Who** should initiate scaffolding? teacher, learner, or system?
2. **When** is the most effective moment to introduce such support?

## 1.4 | Scaffold initiation as a design variable

The decision of when to initiate scaffolding is not pedagogically neutral. Three distinct theoretical traditions converge on this point. From a cognitive load perspective [20], scaffolding that arrives when a learner is already productively engaged risks introducing extraneous processing demands that interrupt rather than sustain ongoing work, the scaffold becomes an obstacle rather than an aid. From a help-seeking perspective [21, 22], learners who request support at a self-detected moment of impasse are cognitively prepared to receive and integrate it; those who receive unsolicited support at an ill-timed moment are not, and may therefore derive little benefit from it or be actively disrupted by it. From an instructional design perspective [3, 23], just-in-time support that is contextually aligned with the learner's current task state is more effective than support that arrives out of step with their needs. These considerations motivate a distinction between three fundamentally different initiation philosophies that this study operationalizes as follows:

- **Teacher-initiated EMME (T-EMME):** support delivered at instructor-defined points.
- **Learner-initiated EMME (L-EMME):** learners request support on demand
- **Automated EMME (Auto-EMME):** support triggered by a physiological indicator of mental effort.

**Teacher-initiated EMME** places initiation control with an instructor who uses pedagogical knowledge to anticipate where learners will struggle. **Learner-initiated EMME** places control with the learner, who initiates support when they judge it necessary. **Automated EMME** places control with a system that uses a real-time physiological indicator, the Index of Pupillary Activity[24], to detect sustained low mental effort and trigger support accordingly. Each mode embeds a different assumption about who can best judge when support is needed, and each carries different risks: teachers may misjudge individual difficulty, learners may fail to recognize their own impasses, and systems may misinterpret physiological signals that are ambiguous in complex task contexts.

## 1.5 | Research aims and hypotheses

This study pursues four interrelated objectives, each operationalized through a testable hypothesis. **The first objective** is to determine whether EMME constitutes an effective scaffold for novice programmers' debugging performance relative to no support. **The second objective** is to examine whether scaffold initiation mode specifically, who controls when EMME is delivered, determines its effectiveness, and whether automated physiological triggering based on a single low-effort threshold operationalized through the Index of Pupillary Activity [24] constitutes a viable basis for adaptive delivery in a complex problem-solving task. **The third objective** is to examine whether EMME functions as a democratizing resource by reducing learners' reliance on prior programming knowledge and attenuating expertise-related performance differences. **The fourth objective** is to characterize productive and disruptive engagement with EMME at the behavioral level by analyzing video interaction logs across conditions. Together, these four objectives advance a unified argument: that the effectiveness of EMME in complex problem-solving depends not on the gaze overlay alone, but on whether its delivery is timed and controlled in ways that align with learners' moment-to-moment needs. These objectives are operationalized through the following hypotheses:

- **H1:** All EMME conditions (T-EMME, L-EMME, Auto-EMME) outperform a no-scaffold control.
- **H2:** Human-mediated initiation (T-EMME, L-EMME) outperforms automated triggering.
- **H3:** Learners with less programming expertise benefit more from EMME scaffolding.
- **H4:** Interaction behaviors with EMME videos (e.g., pausing, replaying) are associated with performance.

## 2 | BACKGROUND

### 2.1 | Effects of EMME on Learner Performance and Processes

Eye Movement Modeling Examples serve as a pedagogical tool designed to scaffold attention by overlaying an expert's gaze onto instructional materials, allowing learners to observe not just task outcomes but the perceptual and cognitive processes that produce them [25]. By directing learners' visual attention toward task-relevant elements, EMME draws on the human propensity for joint attention [26]. and extends it to recorded expert behavior, externalizing covert cognitive strategies that are often absent from traditional verbal explanations [27, 28]. This alignment allows learners to infer task-relevant information more efficiently, as the gaze replay functions as a cognitive scaffold that regulates attentional allocation during complex multimedia tasks [29].

Despite these theoretical advantages, researchers have recently questioned whether the mere observation of a model's gaze inherently translates into improved learning outcomes, as empirical evidence comparing EMMEs to standard video instruction remains inconsistent [30, 31]. Some studies indicate that while EMMEs effectively guide saccadic behavior toward cued regions, this improved synchronization does not reliably yield superior performance on post-test assessments [30, 31]. The discrepancy is most pronounced in cognitively complex and procedural tasks while effects are more reliable in perceptual and search-based tasks with strong visual components [32].

Several moderating factors help explain this variability. Individual differences in prior knowledge and self-regulation capacity shape how successfully learners internalize gaze cues: novices may lack the domain knowledge needed to interpret what the expert's gaze is selecting for, while more advanced learners may find such guidance redundant [29, 27, 31]. The social dimension of human gaze cues implies that learners may perceive and interpret these patterns differently based on their ability to recognize the model's intent, potentially complicating the instructional value of such abstract representations [29]. In procedural tasks, physical interaction demands, such as mouse navigation, may simultaneously distract learners and render gaze-based guidance redundant [33, 34]. At the design level, effectiveness appears to depend on whether EMME emphasizes naturalistic expert behavior or highly curated didactic guidance, and the lack of standardized creation guidelines makes it difficult to isolate which specific design factors drive successful learning across studies [35, 36].

Critically, however, these moderators share a common mechanism that the literature has not yet made explicit: each one affects whether the learner is cognitively positioned to use the scaffold productively at the moment it arrives. A novice who lacks the prerequisite knowledge to interpret expert gaze, a learner whose working memory is already saturated by task demands, and a learner who receives EMME during a moment of productive fluency rather than genuine impasse, all face the same problem: the scaffold is present but not usable [20, 21]. This observation points to scaffold initiation timing and control as a unifying design variable that cuts across the moderators reviewed above. Most prior EMME studies have delivered the intervention at fixed, predetermined points, at the start of a task, between tasks, or as a continuous overlay, without examining how the moment and agent of delivery interact with learners' ongoing cognitive and regulatory states [35, 36]. The present study addresses this gap directly by treating initiation mode as the primary independent variable, testing whether teacher-initiated, learner-initiated, and automated physiological triggering differ in their capacity to deliver EMME at moments when learners can productively integrate it.

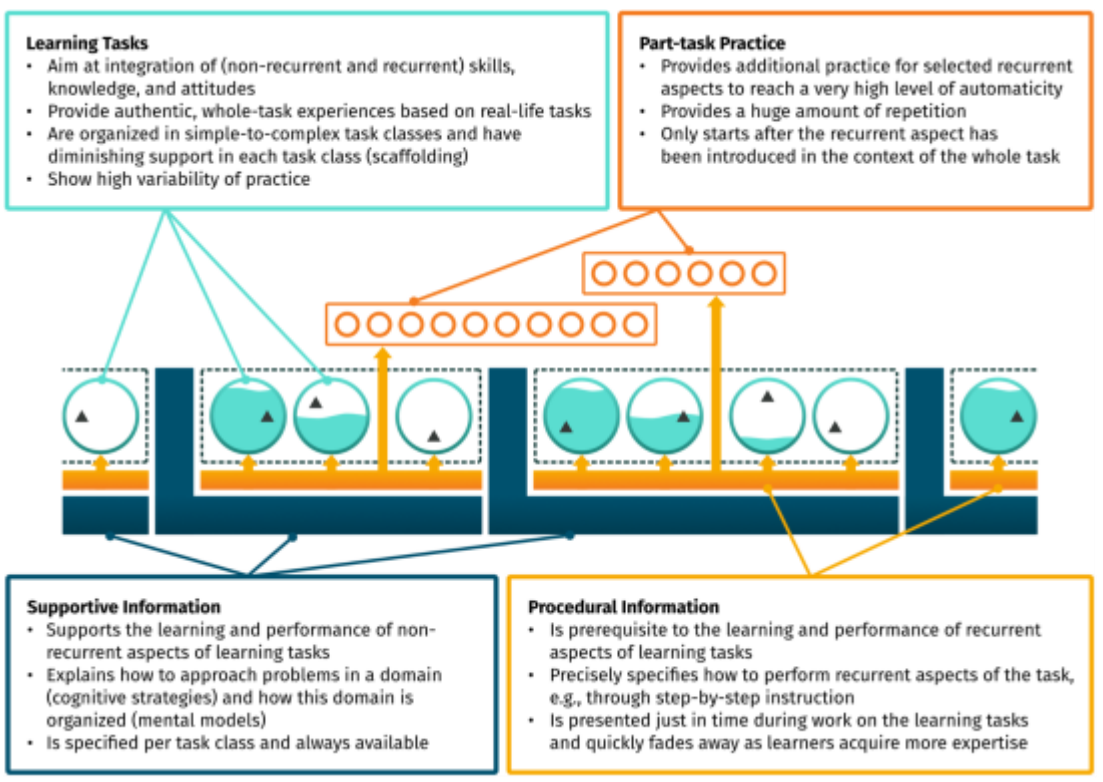

**FIGURE 1** The Four-Component Instructional Design (4C/ID) model, emphasizing whole-task learning and scaffolded skill integration. Adapted from [3].

## 2.2 | EMME in Programming Education

Eye Movement Modeling Examples have emerged as a promising instructional approach in programming education by making otherwise implicit expert problem-solving strategies observable to novices. Empirical evidence suggests that EMME can enhance learners' attention allocation, accelerate time to first fixation on relevant code regions, and improve learning outcomes, particularly for learners with lower prior knowledge [8, 37]. In programming contexts, EMME implementations have ranged from visualizations of expert programmers' eye movements during code comprehension [37] to tutorial videos combining gaze overlays with cursor movements and verbal explanations [35]. These studies report improvements in learners' problem-solving efficiency, debugging performance, and code comprehension, indicating that observing expert gaze can help novices adopt more effective visual search and reasoning strategies.

Systematic review evidence further supports the instructional value of EMME across domains, highlighting generally positive effects on attention guidance and learning outcomes [17]. However, findings also reveal important boundary conditions. While EMME tend to be particularly effective for tasks with strong perceptual components, results are more mixed for complex cognitive or procedural tasks such as programming debugging [31, 8, 19]. These inconsistencies suggest that EMME effectiveness is moderated by task characteristics, learner expertise, and instructional design parameters, factors that interact in ways the field has not yet fully accounted for.

Two moderators are particularly consequential for the present study. First, learner characteristics, especially prior programming knowledge, have been identified as a key moderator of EMME effectiveness. Whereas several studies report robust benefits of eye-movement modeling examples for less experienced learners, other work has demonstrated expertise-reversal patterns, showing that such guidance can become redundant or even detrimental when learners lack the prerequisite knowledge to meaningfully interpret the modeled strategies [38]. This raises the important question of whether EMME can reduce learners' reliance on prior knowledge and support a broader range of learners, rather than amplifying existing performance differences,a question directly addressed by H3 in the present study. Second, interaction behavior with EMME materials appears to play a critical role. Prior work indicates that active engagement behaviors such as transitioning strategically between the instructional video and the task are associated with improved performance [37]. In contrast, excessive pausing or rewinding may signal cognitive overload or misalignment between the provided support and learners' moment-to-moment needs [31, 8]. Despite these insights, little is known about how such interaction behaviors vary across different scaffold initiation modes or whether they mediate the impact of EMME on learning outcomes, a question directly addressed by H4 in the present study.

## 2.3 | EMME and its Empirical and Theoretical Foundation

Eye Movement Modeling Examples (EMME) are instructional videos that overlay an expert's gaze patterns onto a screen recording of their task performance, making otherwise implicit cognitive and attentional strategies explicit to learners [10]. This approach is grounded in several established learning theories, which collectively explain its pedagogical potential.

### 2.3.1 | Observational learning and example-based instruction

EMME is rooted in **Observational Learning Theory** [11], which posits that people learn effectively by observing and imitating a model's behavior. While traditional modeling techniques like cognitive apprenticeship [39] and verbalization protocols [40] externalize verbal reasoning, they fall short for visually complex tasks where knowing *where* to look is crucial [41]. EMME addresses this by visually demonstrating the expert's attentional focus, providing a direct model of the perceptual process. Decades of research have confirmed that studying how a model executes a task, often termed example-based learning or modeling, is more effective than learning through problem-solving alone [42, 43].

### 2.3.2 | Joint attention and gaze following

The mechanism through which EMME guides attention is underpinned by the human propensity for **Joint Attention** [44]. Following another's gaze is an innate human behavior that serves as a fundamental learning mechanism, helping infants learn word meanings and language by following adults' gazes [45, 46]. This mechanism is transferable to eye movement recordings [47], allowing EMME to establish a form of joint attention with the learner. This reduces the need for learners to scan the screen aimlessly, allowing them to focus on integrating visual and auditory information, leading to better comprehension [48, 49].

### 2.3.3 | Cognitive load theory and the signaling principle

From the perspective of **Cognitive Load Theory (CLT)** [20], debugging is a cognitively demanding, ill-structured task that can overwhelm a novice's working memory. EMME helps manage cognitive load [13] by employing the *Signaling*

*Principle* [50]; the gaze overlay acts as a visual cue that directs attention to critical elements, thereby reducing extraneous load caused by disorientation and inefficient scanning. This allows for a greater allocation of mental resources to germane processes, such as building a mental model of the code and the debugging process.

### 2.3.4 | EMME within the 4C/ID framework

This study situates EMME within the **Four-Component Instructional Design (4C/ID) model** [3, 23], a framework for teaching complex skills. The 4C/ID model emphasizes learning through authentic, whole tasks complemented by "just-in-time" supportive information. As illustrated in Figure 1, EMME functions as a concrete form of this supportive information, providing learners with the strategic knowledge and mental models needed to approach non-recurrent aspects of debugging, such as how to systematically form and test hypotheses about errors [51].

In programming education, where tasks like debugging require not only procedural fluency but also the ability to formulate hypotheses, evaluate evidence, and reason through ambiguous code behavior [37, 35], EMME offers a powerful means to bridge the gap between novice and expert problem-solving behaviors [52]. By externalizing an expert's visual and cognitive workflow, EMME can scaffold the acquisition of these complex skills.

### 2.3.5 | Self-Regulated Learning

From a **self-regulated learning (SRL)** perspective, Eye Movement Modeling Examples (EMME) can be understood as ***regulatory resources*** that support learners' monitoring, control, and adaptation during complex problem solving [38]. Debugging is an ill-structured task that requires learners to continuously evaluate their understanding, detect impasses, and decide when and how to regulate their strategies, processes that align closely with cyclical SRL models of monitoring and control [38, 53]. By externalizing expert attentional and strategic processes, EMME provides learners with a metacognitive reference that can support monitoring by revealing mismatches between their own visual search and expert strategies, thereby informing subsequent regulatory decisions.

Critically, the regulatory effectiveness of EMME depends on how *regulatory triggers are detected and enacted*. Learner-initiated EMME aligns with adaptive help-seeking, embedding scaffold use within learners' own regulation cycles. Teacher-initiated EMME represents guided regulation, in which instructors anticipate likely breakdown points that novices may fail to recognize, thereby supporting monitoring without overriding learner agency. System-triggered EMME constitutes a form of system-regulation that relies on inferred indicators of regulatory breakdowns, but its effectiveness depends on alignment with learners' moment-to-moment regulatory needs. Together, these initiation modes reflect distinct forms of regulation self-, guided-, and system-regulated; highlighting scaffold initiation as a central SRL design dimension that can differentially support or disrupt self-regulated debugging in programming education [21, 22, 54].

## 2.4 | Moderators of EMME Effectiveness: Prior Knowledge and Interaction Behavior

Two moderators are particularly relevant to the present study: prior programming knowledge and learners' interaction behavior with EMME materials. Understanding how these variables shape EMME effectiveness is necessary to interpret the hypothesis-driven design of the experiment reported here.

### 2.4.1 | Prior knowledge as a moderator

Research suggests that EMME effectiveness may be moderated by learner characteristics, particularly prior programming knowledge. While some studies report robust benefits for novices [8, 37, 17], others highlight expertise reversal effects, where instructional aids can hinder performance beyond a certain skill threshold [38]. This underscores the need to examine whether EMME can level the playing field by reducing reliance on prior knowledge, thereby supporting a broader range of learners.

### 2.4.2 | Interaction behavior as a moderator

Another critical yet understudied dimension is learner interaction behavior with EMME videos. Prior findings indicate that active engagement, such as transitioning between instructional video and task, can enhance performance [37], while excessive pausing or rewinding may signal cognitive overload or misalignment of support [17, 8]. However, little is known about how these behaviors differ across scaffold initiation modes or whether they mediate the impact of EMME on learning outcomes.

Critically, little is known about how these interaction patterns differ across scaffold initiation modes. The present study addresses this gap directly: by comparing behavioral profiles across T-EMME, L-EMME, and Auto-EMME conditions, it examines whether initiation mode shapes the quality of engagement , productive and integrative versus disruptive and reactive, and whether these behavioral differences mediate the impact of EMME on performance. This analysis forms the empirical basis for H4.

## 2.5 | Additional design and contextual factors moderating EMME effectiveness.

By externalizing covert cognitive strategies, EMMEs visual displays facilitate the acquisition of expert processing heuristics, which are often absent from traditional verbal explanations ([27, 28]. Furthermore, this alignment allows learners to infer task-relevant information more efficiently, as the gaze replay functions as a cognitive scaffold that regulates attentional allocation during complex multimedia tasks [29]. Despite these theoretical advantages, researchers have recently questioned whether the mere observation of a model's gaze inherently translates into improved learning outcomes, as empirical evidence comparing EMMEs to standard video instruction remains inconsistent [30, 31]. Some studies indicate that while EMMEs effectively guide saccadic behavior toward cued regions, this improved synchronization does not reliably yield superior performance on post-test assessments ([30, 31].

**Task complexity.**

This discrepancy is further complicated by **task complexity**, as EMMEs have demonstrated efficacy in perceptual and search-based learning while frequently failing to enhance outcomes in procedural problem-solving [32, 17, 8], Programming debugging exemplifies this challenging profile – it is procedural, distraction-rich, and demands sustained cognitive reasoning rather than perceptual pattern recognition. This pattern underscores why the present study treats debugging as a demanding but important test case for EMME as a scaffold.

**Verbal explanations.**

Moreover, the integration of **verbal explanations** within these displays can sometimes lead to cognitive overload, paradoxically decreasing problem-solving performance compared to non-verbalized modeling [31], which informed the decision in the present study to present EMME without narration.

**Learner self-regulation capacity.**

Individual differences in prior knowledge and cognitive self-regulation capabilities may moderate the extent to which learners successfully internalize these gaze-based cues [29]. For instance, learners with low domain expertise might fail to follow the gaze display or find it redundant when expert verbalizations are already sufficient to focus attention[27, 31].

**Social dimension of human gaze cues.**

Additionally, the **social dimension of human gaze cues** implies that learners may perceive and interpret these patterns differently based on their ability to recognize the model's intent, potentially complicating the instructional value of such abstract representations [29]. Consequently, future investigations should scrutinize task ambiguity and the specific instructional context, as EMMEs may prove particularly advantageous only when verbal explanations are insufficiently precise to guide visual processing [33]. In this regard, researchers must determine whether the physical interaction requirements inherent in procedural tasks, such as mouse navigation, simultaneously distract learners and render gaze-based guidance redundant [33, 34]. Moreover, scholarly discourse suggests that the effectiveness of these modeling techniques may be heavily contingent upon the chosen instructional design, such as whether the EMME emphasizes naturalistic expert behavior versus highly curated didactic guidance [35].

**Lack of standardized creation guidelines.**

The **lack of standardized creation guidelines** remains a significant barrier to understanding, as the wide variability in EMME characteristics across studies complicates the ability to isolate specific design factors that contribute to successful learning [35]. Moreover, it remains critical to distinguish between visual cueing that merely directs attention and interventions that explicitly teach perceptual strategies, as the latter may necessitate different analytical metrics such as minimal distance or entry time to accurately capture learning gains. Indeed, future inquiries must clarify whether such metrics effectively differentiate between simple attentional guidance and the development of higher-order cognitive processing strategies [30]. Additionally, empirical findings suggest that the influence of EMMEs on integrative viewing behavior is inconsistent, raising questions about whether such patterns indicate genuine cognitive integration or merely signify informational overload [32]. Furthermore, the effectiveness of these modeling examples is likely moderated by task expertise, as learners with pre-existing knowledge may derive greater benefit from gaze-based cues when acquiring complex procedural skills.

Beyond these moderators, a further body of work highlights the gap between EMME's theoretical potential and its current empirical realisation. Scholars have proposed that EMMEs might provide more direct access to cognitive processes than traditional verbal interventions, such as strategy worksheets or prompts, by offering visible manifestations of the model's internal decision-making [38]. Additionally, researchers should consider investigating whether these gaze-based cues enhance metacognitive awareness, such as the ability to self-monitor strategy selection during task performance, rather than focusing solely on the direct transfer of perceptual processing patterns [55]. Specifically, evaluating whether these cues successfully prime novice learners to adopt expert-like attentional strategies, even in the absence of explicit verbal instruction, could reveal whether EMMEs fundamentally reshape task-specific visuomotor trajectories [56].

**Taken together, the evidence reviewed above converges on a shared implication:**

the effectiveness of EMME is not determined solely by the gaze overlay itself, but by whether that overlay reaches learners at a moment when they can productively integrate it. Task complexity, the absence of verbal narration, and learner expertise all moderate EMME effectiveness, but these moderators share a common mechanism — they each affect whether the learner is cognitively positioned to use the scaffold when it arrives. This positions scaffold initiation timing and control as a unifying design variable that cuts across these moderators, and provides the conceptual basis for the four hypotheses tested in the present study.

## 2.6 | Methodological Themes in Empirical EMME Research

Empirical research on EMME has employed a range of research designs, and situating the present study within this landscape helps clarify both the gap it addresses and the methodological choices it makes. To achieve this, we reviewed recent empirical contributions using EMME from a research design perspective and identified four emerging themes.

A dominant theme across the dataset is the use of **controlled experimental and randomized designs**, often employing between-subjects or factorial structures (e.g., 2×2, 2×3) [57, 12, 58, 59, 60, 27, 61, 62, 63, 29, 64, 65, 36, 32, 66, 67]. These studies systematically manipulate variables such as EMME presence, prior knowledge, cueing type, or instructional format to establish causal effects on learning outcomes. Many adopt randomized controlled trials (RCTs) or tightly controlled lab experiments, frequently integrating eye-tracking measures to capture cognitive processes alongside performance. This reflects a strong emphasis in the field on internal validity and causal inference, positioning EMME as an intervention tested under rigorous experimental conditions.

Another group of studies uses **pre–post or quasi-experimental designs**, often in more applied or educational settings [25, 28, 35, 68, 37, 69]. These designs assess learning gains over time, typically comparing performance before and after exposure to EMME-based instruction. While less tightly controlled than RCTs, they provide ecologically valid insights into instructional effectiveness, especially in classroom or training contexts. These studies often focus on intervention impact, highlighting how EMME influences learning processes longitudinally rather than isolating variables in controlled environments.

A large portion of the dataset consists of **eye-tracking-based, comparative, or exploratory studies** [70, 13, 56, 71, 72, 9, 73, 74, 75, 31, 76, 77, 78, 79, 80, 81, 82, 82, 34, 83, 26, 84, 38, 15, 85, 33, 55]. These works focus on analyzing visual attention, gaze patterns, and cognitive processing rather than strictly testing interventions. Many compare expert vs novice behavior, different cueing methods (e.g., gaze vs cursor), or variations of EMME implementation. This theme reflects the methodological core of the field, where eye-tracking is used both as a measurement tool and as an instructional mechanism, bridging cognitive science and educational research. These studies contribute to understanding how EMME works, not just whether it works.

A smaller but important theme includes **conceptual or theoretical papers** [30], which do not involve empirical participants but instead propose frameworks, design principles, or pedagogical approaches related to EMME. These works synthesize existing evidence and extend theoretical understanding, often introducing new constructs or instructional paradigms. Although limited in number, they play a key role in guiding future empirical research and interpreting findings across studies.

## 2.7 | A Regulatory Turn: Foregrounding Scaffold Initiation and Learner Agency

Our experiment introduces a notable shift from the dominant themes by foregrounding **scaffold initiation** and **learner regulation** as the central design variable, rather than focusing primarily on instructional formats, cueing variations, or comparative effectiveness of EMME. While it still adopts a **controlled between-subjects experimental design** (aligning with Theme 1), its contribution goes beyond typical causal testing by **reframing EMME as a regulatory scaffold embedded within self-regulated learning (SRL) processes**, explicitly examining who controls support and when it is delivered.

Unlike most experimental studies that treat EMME as a static intervention, and unlike exploratory eye-tracking studies that focus on attention patterns, this work integrates **process-level behavioral analytics** (e.g., transitions, pauses, navigation patterns) to link scaffolding modes with regulatory engagement. It also diverges from quasi-experimental classroom studies by not primarily assessing learning gains over time, but instead analyzing the alignment (or misalignment) between adaptive system triggers and learners' cognitive-regulatory states.

Critically, our experiment challenges a common assumption in the field, that more automation improves adaptivity, by showing that **fully automated, data-driven scaffolding can disrupt learning** when it overrides learner agency, whereas human-mediated or learner-controlled support enhances performance and regulation. In doing so, it advances the literature by introducing a conceptual distinction between self-, guided-, and system-regulated scaffolding, positioning EMME not merely as an instructional tool, but as a mechanism **whose effectiveness depends on its integration within learners' regulatory cycles rather than its mere presence**.

# 3 | METHODOLOGY

This section describes the experimental design, participants, learning environment, EMME materials, scaffold initiation conditions, procedure, and measurement framework used to address the four previously outlined hypotheses.

## 3.1 | Experimental Design and Participants

We employed a between-subjects design with four conditions: Teacher-Initiated EMME (T-EMME), Learner-Initiated EMME (L-EMME), Automated EMME (Auto-EMME), and a no-scaffold control. Participants were randomly assigned to conditions, ensuring that individual differences in prior knowledge, motivation, and programming ability were distributed across groups by chance rather than by design.

A total of 120 undergraduate computer science students from a European university participated (31 female; mean age = 20.78 years, SD = 2.73). Inclusion required completion of at least one introductory programming course, ensuring participants possessed foundational Python knowledge while remaining representative of the novice programmer population the intervention was designed to support. Thirty participants were randomly assigned to each of the four conditions, yielding a balanced design. This sample size provided sufficient statistical power to detect meaningful differences while maintaining practical feasibility for the intensive eye-tracking and interaction data collection.

## 3.2 | The Learning Environment

We developed a custom instructional environment integrating Eye Movement Modeling Examples to support novice programmers during Python debugging tasks. As shown in Figure 2, the interface comprised two panels: a main panel displaying the Python debugging task, and a secondary panel in the right corner hosting the EMME video player. The main panel presented the complete program source code, which participants were required to debug.

The EMME video player displayed the expert's screen recording with a gaze overlay rendered as a moving dot indicating the expert's current and recent fixation locations. To preserve learner autonomy across all EMME conditions, a 'Disable Help' button allowed participants to temporarily hide the EMME panel. Video controls mirrored professional media players: play/pause, ±5-second navigation jumps, playback speed adjustment (0.5× to 2.0×), volume control, and interactive timeline seeking.

```
1
2  import pygame
3
4  BLACK = (0, 0, 0)
5  WHITE = (255, 255, 255)
6  BLUE = (0, 0, 255)
7  GREEN = (0, 255, 0)
8  RED = (255, 0, 0)
9  PURPLE = (255, 0, 255)
10
11
12 class Wall(pygame.sprite.Sprite):
13     """This class represents the bar at the bottom that the player controls """
14
15     def __init__(self, x, y, width, height, color):
16         """ Constructor function """
17
18         # Call the parent's constructor
19         super().__init__()
20
21         # Make a BLUE wall, of the size specified in the parameters
22         self.image = pygame.Surface([width, height])
23         self.image.fill(color)
24
25         # Make our top-left corner the passed-in location.
26         self.rect = self.image.get_rect()
27         self.rect.y = y
28         self.rect.x = x
29
30
31 class Player(pygame.sprite.Sprite):
32     """ This class represents the bar at the bottom that the
33     player controls """
34
35     # Set speed vector
36     change_x = 0
```

EMME Feedback

[Speed: 1.0×] [Mute] [Disable Help ✕]

Ln 1, Col 1 Spaces: 4 UTF-8 CRLF Python Go Live

**FIGURE 2** Example of EMME feedback demonstrated in Visual Studio

## 3.3 | EMME Materials

### 3.3.1 | Expert selection and video production

The EMME videos were produced through an iterative development process. Our initial attempts to record programming teachers revealed an interesting phenomenon: their expert efficiency actually worked against effective modeling. Teachers solved problems so rapidly that their gaze patterns became compressed and difficult to interpret. This led us to pivot toward teaching assistants, whose more deliberate problem-solving pace and explicit reasoning processes created richer, more informative EMME content. The teaching assistants' longer debugging sessions, complete with visible hypothesis testing and debugging, provided novice learners with a more realistic and accessible model of the debugging process. Therefore, the final EMME video featured this teaching assistant's screen recordings and gaze data in absence of verbal explanation to evade cognitive overload

### 3.3.2 | Composition of EMME Learning Materials

Each Eye Movement Modeling Example combined two carefully synchronized components that worked in concert to externalize expert cognition:

**Expert's Screen Recording:** These videos captured the complete debugging journey of teaching assistants, from initial code reading through final solution. Rather than presenting polished, linear solutions, we preserved the authentic problem-solving process including moments of confusion, exploratory testing, and iterative refinement. This raw footage demonstrated not just what experts do, but how they think when confronted with ambiguous code behavior. Therefore, through this recording, participants could observe how the expert behaves during code reading and analysis, scrolling through the program, inserting diagnostic print statements, and modifying code to fix logical bugs.
**Expert´s Gaze Visualization:** Superimposed on the screen recording, we displayed a dynamic representation of the expert's visual attention. Using eye-tracking data, we created the signal as a translucent white dot that indicated the expert's current focus point, accompanied by a persistent trail that revealed their scanning patterns over time. This visualization transformed abstract cognitive processes into concrete visual cues, allowing novices to literally "see" how experts systematically examine code, identify suspicious patterns, and verify hypotheses.

## 3.4 | Scaffold Initiation Conditions

We implemented three fundamentally different philosophies about when and how debugging support should be provided:

### Teacher Initiated EMME (T-EMME): The Guided Approach

In this condition, scaffolding appeared automatically for problems the instructor had pre-identified as particularly challenging. This approach embodied the traditional educational philosophy of expert-curated support–the teacher anticipating where students would struggle and providing guidance precisely at those moments. Participants retained agency through the ability to disable the help, but the default presentation ensured they encountered expert strategies for the most conceptually difficult problems.

### Learner Initiated EMME (L-EMME)

This condition placed control firmly in learners' hands through a prominent "Help Me" button. Participants decided for themselves when they needed support, reflecting principles of self-regulated learning and metacognitive awareness. This approach respected learners' ability to monitor their own understanding and request assistance only when they recognized their own knowledge gaps or reached impasses in their problem-solving process.

### Automated EMME (Auto-EMME)

The Auto-EMME condition operationalized the IPA-based cognitive load measure [24] described in Section 3.6.3 to deliver support automatically, without any instructor or learner initiation. Using real-time eye-tracking data, the system continuously monitored each participant's cognitive engagement and activated scaffolding support when their IPA remained more than two standard deviations below their individual resting-state baseline for a continuous period exceeding 60 seconds. This threshold-based mechanism ensured that support was triggered only when a meaningful and sustained decline in cognitive engagement was detected, rather than in response to transient fluctuations. In this way, the Auto-EMME condition provided an adaptive, data-driven alternative to the teacher- and learner-initiated support conditions.

## 3.5 | Procedure

The study unfolded through a carefully orchestrated sequence designed to balance experimental control with ecological validity:
The session began with informed consent, ensuring participants understood their rights and the study's purpose. Participants then completed a Python programming pre-test, establishing their baseline expertise level. Following this assessment, we provided a comprehensive orientation to the learning environment, including hands-on practice with all interface features and EMME controls.

We then calibrated the Tobii-X120 eye-tracker for each participant, ensuring precise measurement of visual attention and cognitive load throughout the debugging tasks. The core of the session involved solving ten Python programs, each containing logical bugs that required systematic analysis and correction. We explicitly informed participants that the programs contained no syntax errors, focusing their attention exclusively on logical reasoning and program behavior.

## 3.6 | Measures

### 3.6.1 | Debugging Performance

We measured debugging success through a straightforward but meaningful metric: the number of programs correctly debugged from the set of ten. Each successful solution earned one point, creating a performance scale from 0 to 10 that reflected participants' practical problem-solving ability in identifying and resolving logical errors in Python code.

### 3.6.2 | Prior programming expertise

The Python pre-test provided a robust measure of participants' prior programming knowledge, serving as both a screening tool and a statistical covariate. This 10-point assessment evaluated fundamental programming concepts relevant to the debugging tasks, allowing us to account for individual differences in starting competence.

### 3.6.3 | Cognitive load (IPA)

Cognitive load was assessed continuously throughout the debugging task using the Index of Pupillary Activity (IPA) [24], a psychophysiological measure derived from real-time eye-tracking data. The IPA captures moment-to-moment fluctuations in pupil dilation that closely reflect mental effort, making it well-suited for monitoring the dynamic cognitive demands of programming tasks in an unobtrusive manner. To account for inter-individual variability in baseline pupil size, each participant's resting-state IPA was recorded during a quiet period prior to the task. This personal baseline served as the reference against which task-period IPA values were continuously compared. A sustained depression in IPA – defined as values falling more than two standard deviations below the individual's baseline for longer than 60 consecutive seconds – was interpreted as an indicator of disengagement or insufficient cognitive engagement. As described in the Automated EMME condition below, this threshold formed the core trigger criterion for adaptive support delivery.

### 3.6.4 | Behavioral Engagement

Beyond simple performance metrics, we captured rich interaction data that revealed how participants engaged with the EMME support system. In the experiment, the participants were given the full control of the EMME videos. They can not only choose to play/pause, move back/forward in the video but they could also select how much eye-tracking data should be overlayed on the video. Their choice range was from one second to the entire length of the video. To adhere to this, we define the following variables capturing participants' video watching behaviour.

- Number of pauses: total number of pauses during the session.
- Average duration of pauses: average duration of pauses in seconds.

- Ratio of video duration played: total number of unique seconds played by the participants divided by the total duration of the video in seconds.
- Number of backward jumps: total number of backward jumps during the session.
- Number of forward jumps: total number of forward jumps during the session.
- Average gaze-duration overlayed (scanpath length): this is the average length of the expert's gaze that the participants chose to see on the EMME video.
- Transitions between task and EMME

These behavioral fingerprints provided unprecedented insight into how different scaffolding approaches influenced learning processes and engagement patterns. This comprehensive measurement approach enabled us to move beyond the simple question of "what worked" to explore the more nuanced question of "how different scaffolding strategies influenced learning behaviors and outcomes."

## 4 | RESULTS

This section presents the findings on debugging performance across the four experimental conditions, the moderating role of prior programming expertise, and the relationship between interaction behaviors and performance outcomes.

### 4.1 | Debugging Performance Across Scaffold Conditions

A one-way ANOVA confirmed a significant main effect of the help condition on debugging performance, ($F[3,116] = 132.44$, $p < .05$). As hypothesized (H1), all three EMME conditions significantly outperformed the control condition (see Figure 3 and Table 1 for descriptive statistics). Post-hoc comparisons revealed a clear performance hierarchy, supporting H2. Both the Teacher-Initiated (T-EMME: $M = 6.56$, $SD = 1.16$) and Learner-Initiated (L-EMME: $M = 6.93$, $SD = 0.83$) conditions produced the highest performance levels, with no statistically significant difference between them ($F(1,58) = 1.97$, $p > .05$). The Automated EMME condition (Auto-EMME: $M = 4.06$, $SD = 0.83$) significantly underperformed compared to both T-EMME and L-EMME but still surpassed the control group (Control: $M = 2.50$, $SD = 1.13$).

### 4.2 | The Moderating Role of Prior Programming Expertise

Across all participants, a modest positive correlation existed between prior programming expertise and debugging performance ($r(118) = 0.20$, $p < .05$). However, this relationship was strongly conditioned (H3). A strong, significant correlation was found only in the control group ($r = 0.58$, $p < .05$) , whereas in all three EMME conditions, the correlation between expertise and performance was non-significant (Table 3).

An ANCOVA (Model 1, Table 4) confirmed significant main effects for both condition and expertise. More importantly, introducing interaction terms between condition and expertise (Model 2, Table 5) significantly improved the model fit, $F(3, 112) = 4.64$, $p = .004$. The analysis revealed that the positive effect of expertise on performance was significantly reduced in the T-EMME ($\beta = -0.40$, $p = .006$) and Auto-EMME ($\beta = -0.48$, $p = .007$) conditions, with a similar, though non-significant, trend in L-EMME ($\beta = -0.22$, $p = .10$). These results indicate that EMME scaffolding reduced reliance on prior expertise, particularly in the teacher-initiated and automated modes (see Figure 5).

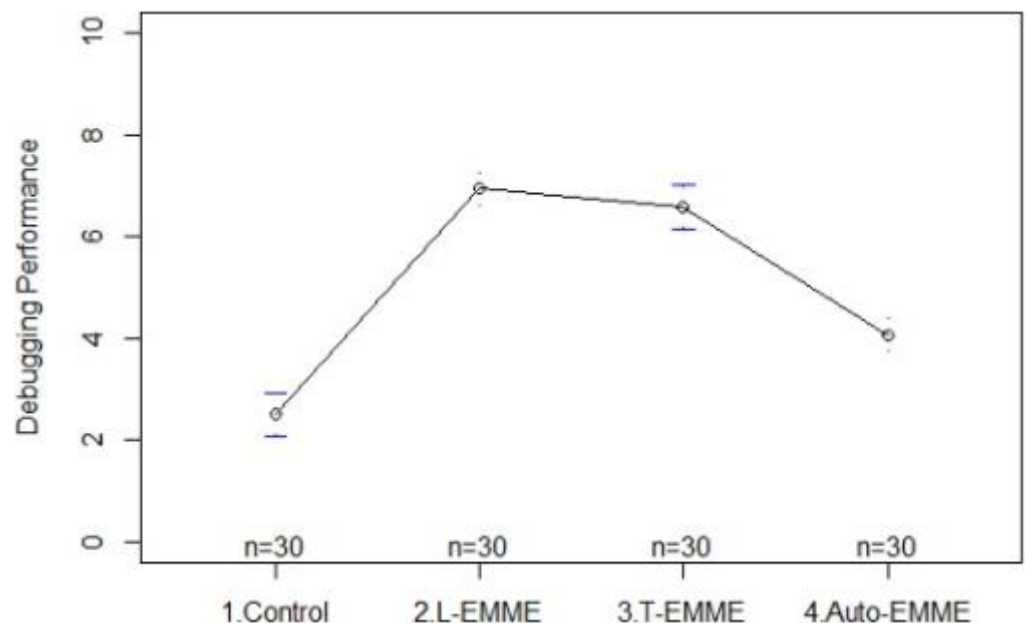

**FIGURE 3** Debugging performance across the four help conditions. The blue bars are the 95% confidence interval

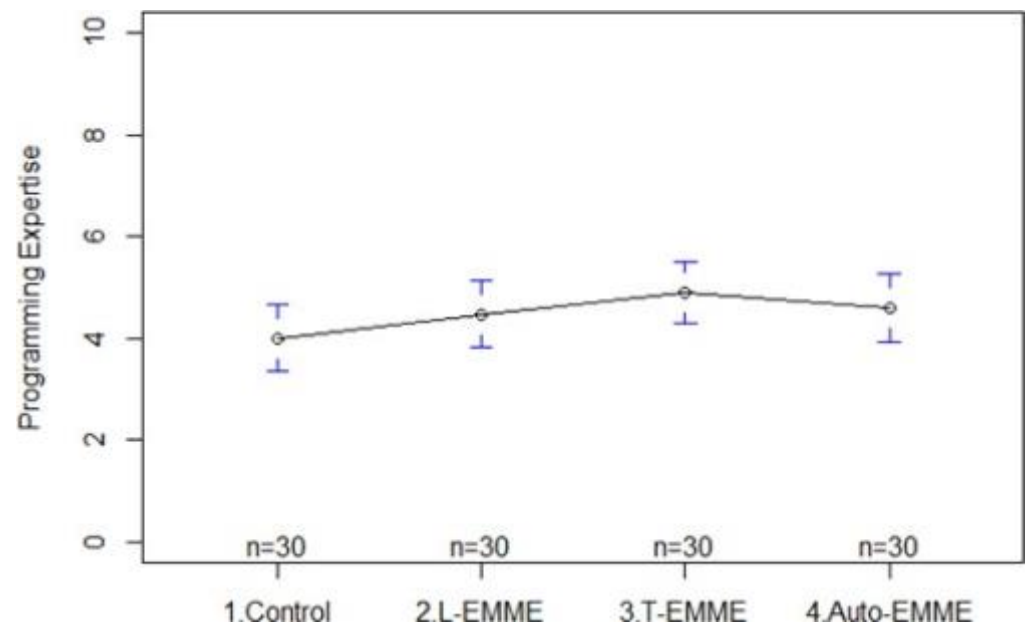

**FIGURE 4** Debugging performance across the four help conditions. The blue bars are the 95% confidence interval

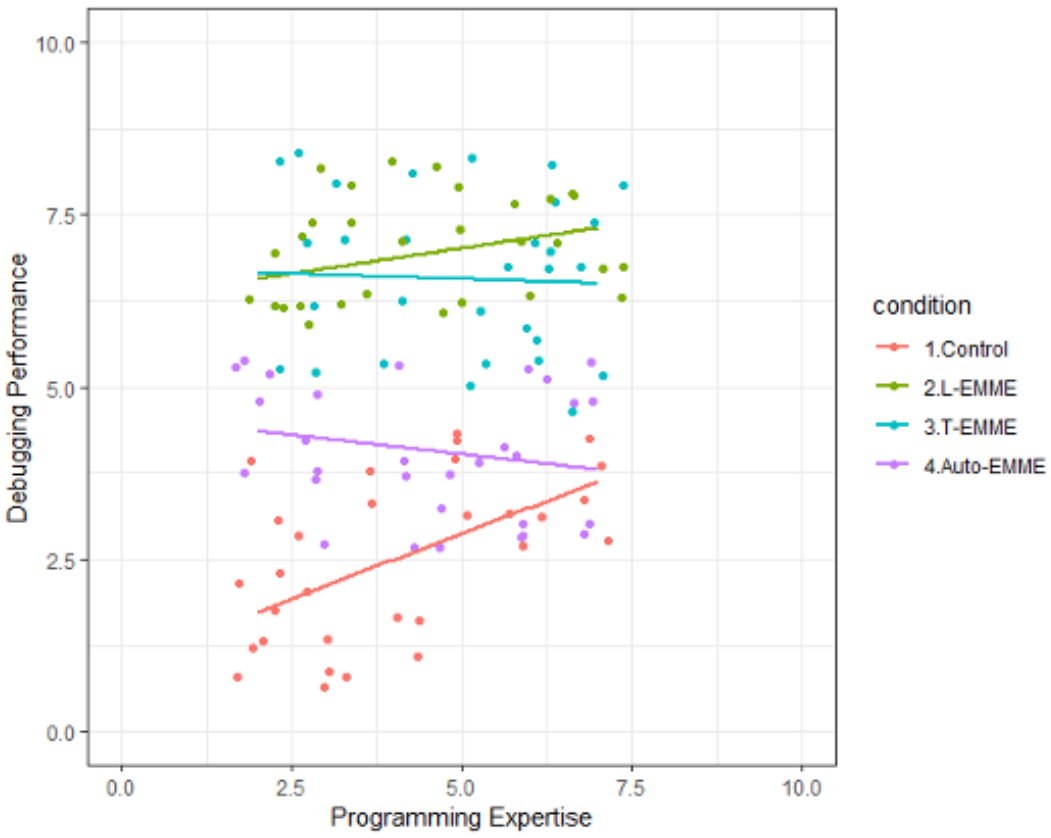

**FIGURE 5** Debugging performance and programming expertise across the four help conditions.

**TABLE 1** Pairwise comparison between the help conditions

| | Control | T-EMME | L-EMME |
|---|---|---|---|
| T-EMME | F[1,58] = 187.18, **p < .05** | | |
| L-EMME | F[1,58] = 298.07, **p < .05** | F[1,58] = 1.97, p > .05 | |
| Auto-EMME | F[1,58] = 37.22, **p < .05** | F[1,58] = 91.79, **p < .05** | F[1,58] = 179.94, **p < .05** |

We observe a significant covariate effect of expertise. However, the covariate effect disappears when we only consider the three EMME conditions.

## 4.3 | EMME Interaction Behaviors and Performance

Analysis of video interaction logs revealed distinct behavioral patterns that correlated with debugging performance, offering partial support for H4. Of the seven behavioral variables recorded, only three yielded significant relationships with performance (Figure 6), and none showed a significant relationship with prior expertise (Table 6). This pattern is itself theoretically informative and is discussed in detail in Section 5.4.

Notably, the four variables that showed no significant relationship – pause duration, video playback proportion, number of forward jumps, and scanpath length – all share a common characteristic: they capture the quantity or duration of video consumption rather than the directionality or strategic patterning of engagement. The three variables that did predict performance – pause frequency, backward jumps, and task-video transitions – capture how learners engaged with the EMME material, not simply how much. This quality-over-quantity pattern provides the empirical basis for interpreting behavioral engagement as strategic rather than consumptive.

- **Negative Engagement:**
  - **Video Pauses:** There was a significantly negative correlation between number of pauses and debugging performance (r(88) = -0.41, p < .00001) – meaning students who paused more tended to perform worse (Figures 6a). There was also a significant difference in pause frequency across the three EMME conditions, with Auto-EMME students pausing significantly more than L-EMME students.

**TABLE 2** Descriptive statistics

| Condition | Mean performance (out of 10) | SD performance | Mean expertise (out of 10) | SD expertise |
|---|---|---|---|---|
| Control | 2.50 | 1.13 | 4.00 | 1.76 |
| T-EMME | 6.56 | 1.16 | 4.90 | 1.60 |
| L-EMME | **6.93** | **0.83** | 4.46 | 1.73 |
| Auto-EMME | 4.06 | 0.83 | 4.60 | 1.77 |

**TABLE 3** Correlation between programming expertise and debugging performance

| Condition | Pearson Correlation | P value |
|---|---|---|
| Control | 0.58 | **< .05** |
| T-EMME | -0.04 | > .05 |
| L-EMME | 0.31 | > .05 |
| Auto-EMME | -0.23 | > .05 |

- **Backward Jumps:** Similarly, there was a significantly negative correlation between the number of backward jumps and debugging performance ($r(88) = -0.52$, $p < .00001$) (Figures 6b) – the strongest correlation among the video engagement variables. Auto-EMME students also jumped backward significantly more than L-EMME students. Crucially, unlike pauses, backward jumps did show a significant covariate interaction with condition (ANOVA model comparison $p = .02$). Specifically, in the T-EMME condition, more backward jumps were especially strongly associated with worse performance, suggesting that repeatedly rewinding the teacher-initiated video was a particularly strong signal of struggle in that condition.

- **Positive Engagement:** Conversely, the **number of transitions** between the EMME video and the debugging task was positively correlated with performance ($r = 0.32$, $p = .002$, Figure 6c). Meaning students who switched back and forth between the EMME video and the task more frequently actually tended to perform better. Moreover, Participants in the L-EMME and T-EMME conditions made significantly more of these transitions than those in the Auto-EMME condition (Table 7).

  - **Differences Across Condition.** There was a significant difference in transition counts across the three EMME conditions ($F[2,87] = 4.95$, $p = .009$). Pairwise comparisons showed that Auto-EMME students made significantly fewer transitions than both L-EMME and T-EMME students.

Further analysis confirmed that the negative relationship between backward jumps and performance was a significant covariate, with The interaction between backward jumps and T-EMME condition was significant (estimate = -0.112, $p = .0008$). This suggests that even in effective conditions, certain disoriented behaviors can be detrimental. The remaining variables showed no significant relationship with either performance or expertise.

**TABLE 4** Model 1: performance ~ condition + expertise

| | Estimate | Std. err. | t-val | p-val |
|---|---|---|---|---|
| Condition (control) | 2.09 | 0.28 | 7.48 | < 0.00001 |
| Condition (L-EMME) | 6.48 | 0.29 | 21.62 | < 0.00001 |
| Condition (T-EMME) | 6.07 | 0.31 | 19.06 | < 0.00001 |
| Condition (Auto-EMME) | 3.60 | 0.30 | 11.79 | < 0.00001 |
| Expertise | 0.10 | 0.05 | 1.87 | 0.006 |

**TABLE 5** Model 2: performance ~ condition * expertise

| | Estimate | Std. err. | t-val | p-val |
|---|---|---|---|---|
| Condition (control) | 0.98 | 0.43 | 2.27 | 0.02 |
| Condition (L-EMME) | 6.27 | 0.48 | 12.94 | < 0.00001 |
| Condition (T-EMME) | 6.71 | 0.56 | 11.89 | < 0.00001 |
| Condition (Auto-EMME) | 4.58 | 0.48 | 9.38 | < 0.00001 |
| Expertise | **0.37** | 0.09 | 3.78 | **0.0002** |
| Condition (L-EMME): expertise | -0.22 | 0.14 | -1.61 | 0.10 |
| Condition (T-EMME): expertise | **-0.40** | 0.14 | -2.75 | **0.006** |
| Condition (Auto-EMME): expertise | **-0.48** | 0.14 | -3.47 | **0.007** |

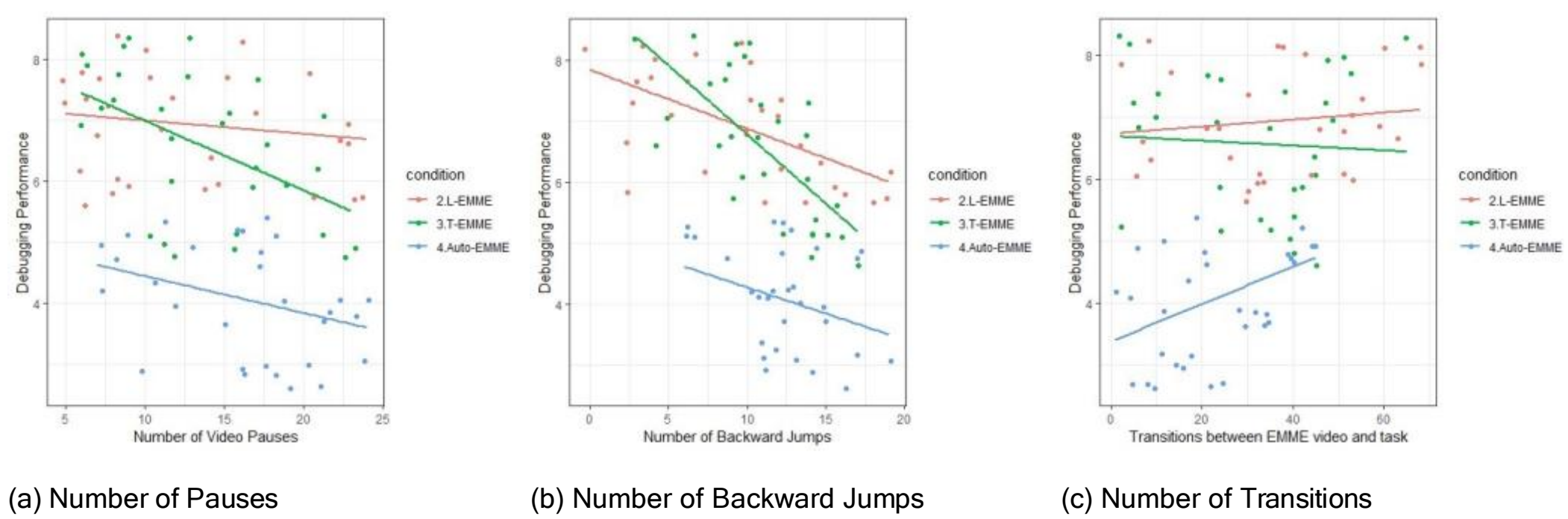

(a) Number of Pauses (b) Number of Backward Jumps (c) Number of Transitions

**FIGURE 6** EMME Interaction Behaviors and Their Relationship to Performance

## 5 | DISCUSSION

This study examined whether scaffold initiation mode, who controls the timing of EMME delivery, determines its effectiveness in supporting novice programmers' debugging performance. The findings provide strong support for H1, H2, and H3, and partial but theoretically patterned support for H4. Together they establish scaffold initiation timing and control as a critical and previously unexamined design variable for EMME, and demonstrate that automated triggering based on a single low-effort physiological threshold is insufficient and potentially counterproductive in complex problem-solving contexts. The discussion addresses each hypothesis in turn before offering an integrated interpretation.

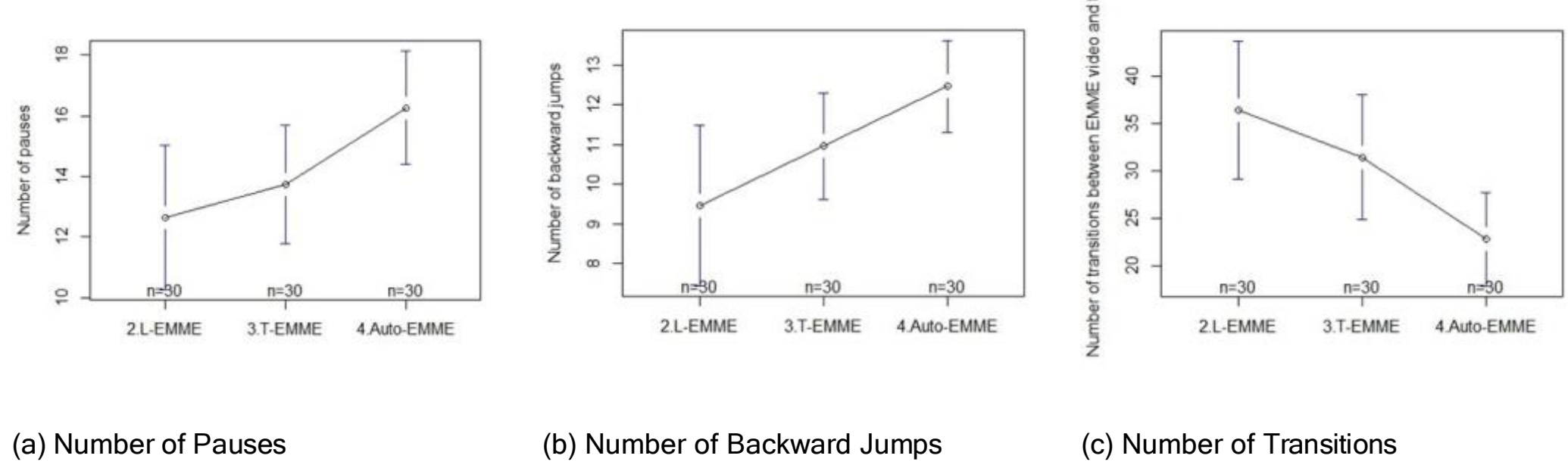

(a) Number of Pauses (b) Number of Backward Jumps (c) Number of Transitions

**FIGURE 7** EMME Interaction Behaviors and Their Relationship to Performance

**TABLE 6** Correlation with expertise and performance

| | Expertise | | Performance | |
|---|---|---|---|---|
| | R | P | R | P |
| Video Pauses | 0.09 | 0.35 | **-0.41** | **<0.00001** |
| Pause duration | -0.08 | 0.40 | -0.17 | 0.10 |
| Video playback proportion | 0.01 | 0.91 | -0.02 | 0.84 |
| Jumps backward | -0.11 | 0.28 | **-0.52** | <**0.00001** |
| Jumps Forward | -0.07 | 0.50 | -0.15 | 0.15 |
| Scanpath length | -0.15 | 0.14 | 0.02 | 0.82 |
| Transitions between tasks and EMME | 0.01 | 0.89 | **0.32** | **0.002** |

## 5.1 | EMME Outperforms No Scaffolding Across All Initiation Modes (H1)

As predicted by **H1**, all EMME conditions led to significantly better debugging performance compared to the no-scaffold control. This finding validates EMME as an effective form of supportive information, operationalizing the 4C/ID model by externalizing expert problem-solving strategies during whole-task learning [3, 23]. This outcome helps to resolve the "mixed findings" in the EMME literature. Aligning with the recent meta-analysis by Tunga and Cagiltay [9],which established a significant, small-to-medium overall effect (d = 0.43), our study demonstrates that EMME reliably enhances performance when cognitive load and task complexity are properly managed. Furthermore, when viewed through the lens of Kalyuga’s expertise reversal framework [86], the so-called "mixed effects" of EMME appear to arise not from the technique itself, but from scaffolds that are poorly timed relative to learners’ cognitive needs. Our findings confirm that variability in EMME's effectiveness is primarily a function of design, not mechanism. By identifying scaffold initiation timing and control as critical factors, we propose that teacher- and learner-initiated modes are key to unlocking EMME's full potential in complex, ill-structured domains like debugging.

**TABLE 7** ANOVA and Pairwise Comparisons Across EMME Conditions

| | ANOVA with 3 feedback types | | L-EMME vs T-EMME | | L-EMME vs Auto-EMME | | T-EMME vs Auto-EMME | |
|---|---|---|---|---|---|---|---|---|
| | F | P | F | P | F | P | F | P |
| Video Pauses | **3.34** | **0.04** | 0.53 | 0.46 | 5.99 | **0.01** | 3.64 | **0.06** |
| Pause Duration | 2.15 | 0.12 | | | | | | |
| Video Playback Proportion | 0.60 | 0.55 | | | | | | |
| Jumps Backward | **3.88** | **0.02** | 1.58 | 0.21 | 6.87 | **0.01** | 3.0 | **0.08** |
| Jumps Forward | 0.79 | 0.45 | | | | | | |
| Scanpath Length | 0.13 | 0.87 | | | | | | |
| Transitions Between Task and EMME | **4.95** | **0.009** | 1.07 | 0.30 | 10.01 | **0.002** | 4.65 | **0.03** |

## 5.2 | Human-Mediated Initiation Outperforms Automated Triggering (H2)

**Teacher- and learner-initiated scaffolding (T/L-EMME) outperformed automated support (Auto-EMME), highlighting the critical importance of human-judged timing and control in delivering instructional support.** Supporting **H2**, both teacher- and learner-initiated scaffolding (T-EMME and L-EMME) were significantly more effective than the automated approach (Auto-EMME). This outcome underscores the critical importance of timing and control in instructional support. The superior performance of T-EMME and L-EMME aligns with principles of Cognitive Load Theory (CLT) [20]; these modes likely delivered support at more cognitively appropriate moments, minimizing extraneous load. In contrast, the Auto-EMME condition, triggered by a physiological measure of low mental effort, may have interrupted productive cognitive engagement, thereby increasing extraneous load and disrupting the learning process. This is consistent with the 4C/ID emphasis [3, 23] on "just-in-time" information, suggesting that system-triggered support requires more nuanced algorithms that consider the learner's current task context, not just a single cognitive metric.

## 5.3 | EMME as an Equalising Resource: Reducing Reliance on Prior Expertise (H3)

**EMME scaffolding functioned as a democratizing tool, leveling the performance field by reducing the advantage associated with prior programming knowledge.** Confirming **H3,** the strong correlation between prior expertise and performance observed in the control group disappeared in all EMME conditions. This leveling effect suggests that EMME can effectively reduce the performance advantage typically associated with prior knowledge, likely by providing all learners with direct access to expert visual and cognitive strategies that they lack.

This finding informs the ongoing debate about prior knowledge and EMME effectiveness[17, 8]. While an expertise reversal effect can sometimes render instructional aids less effective for higher-knowledge learners [30, 26, 87, 32], our results underscore that EMMEs offer significant benefits for novices. By externalizing an expert's visual and cognitive strategies, EMME scaffolds bridge a critical gap for novice programmers, equipping them with strategic approaches that are otherwise only gained through extensive experience. This democratizing potential, that is reducing the performance disparity linked to prior knowledge, is a key contribution to promoting equity in computer science education.

## 5.4 | Behavioral Markers of Productive and Disrupted Regulation (H4)

The analysis of interaction behaviors provided partial support for H4 and revealed a pattern of significant and null results that, while not fully anticipated prior to data collection, is theoretically coherent and carries implications for how future studies should design and interpret behavioral engagement batteries.

"Task-video transitions" were positively correlated with performance ($r = 0.32$, $p = .002$), consistent with an adaptive help-seeking interpretation: learners who alternated purposefully between the expert model and their own debugging task were better able to integrate scaffolded guidance into their ongoing problem-solving. From a help-seeking perspective , each transition represents a complete regulatory cycle: detecting a need, consulting an external resource, and returning to apply what was observed. This behavior was significantly more prevalent in L-EMME and T-EMME than in Auto-EMME, suggesting that human-mediated initiation created conditions more conducive to this integrative engagement pattern.

Conversely, pauses and backward jumps were negatively correlated with performance ($r = -0.41$ and $r = -0.52$, respectively) and were significantly more frequent in Auto-EMME. These variables capture discrete behavioral events with an interpretable directional meaning: pauses represent interruptions to the flow of engagement, and backward jumps represent reversals of the temporal progression through the expert model, both consistent with a learner who has been disoriented by support that arrived at a poorly aligned moment rather than at a genuine point of need. The covariate analysis further showed that the negative relationship between backward jumps and performance was especially pronounced in the T-EMME condition (estimate = -0.112, $p = .0008$), indicating that disoriented navigation behavior can undermine learning even within an otherwise effective scaffolding condition, and suggesting that backward jumping reflects a genuine failure of integration that is not limited to the automated condition.

Collectively, these findings suggest that what distinguishes effective from disrupted scaffolding is not how much learners engage with EMME, but whether their interaction is strategic and integrative or reactive and disoriented. We nonetheless acknowledge that the behavioral variables used here are proxies rather than direct measures of regulatory processes. Firmer conclusions about the underlying monitoring judgments and help-seeking decisions would require complementary process measures such as think-aloud protocols or fine-grained event-level logs. These remain important directions for future research.

# Limitation and Future Research

Several limitations of this study warrant consideration when interpreting the results. First, the automated scaffolding (Auto-EMME) was triggered based on a single psychophysiological measure, that is prolonged low mental effort as indicated by the Index of Pupillary Activity. This approach differs from the other practice of triggering support during both *high* and *low* cognitive load or stress, as seen in studies like [88], where feedback was activated when cognitive load or stress exceeded two standard deviations above a resting baseline. Our decision to trigger support during low mental effort was based on the hypothesis that it might signal disengagement or confusion; however, this appears to have led to mistimed interventions that disrupted productive struggling rather than supporting it. The behavioral data, showing more pauses, backward jumps, and fewer task, video transitions in the Auto-EMME condition support this interpretation.

Second, the reliance on a single modality (eye-tracking) for automated triggering represents a technical limitation. Future systems could benefit from multi-modal approaches that combine eye-tracking with other physiological measures (e.g., heart rate variability for stress) and behavioral indicators (e.g., prolonged inactivity or error patterns) to create more robust and context-sensitive scaffolding algorithms.

Third, the study was conducted in a controlled laboratory setting. While this allowed for rigorous measurement of eye movements and interaction behaviors, it may limit the ecological validity of the findings. Research is needed to examine the long-term effects of these scaffolding methods in authentic classroom environments, where distractions, collaboration, and extended task periods may influence how learners engage with and benefit from EMME.

Finally, while the expert EMME videos were designed to model authentic debugging processes, the content was fixed and not personalized to individual learners' code states or error types. Future systems could explore dynamic EMME content that adapts to the learner's current code context or specific type of error, potentially increasing the relevance and effectiveness of the support.

Based on these limitations and our findings, we propose several directions for future research:

- **Compare triggering approaches**: Future studies should directly compare the effectiveness of scaffolding triggered during high versus low cognitive load states to determine optimal intervention timing for complex problem-solving tasks.
- **Develop multi-modal algorithms**: Explore the integration of multiple data streams (eye-tracking, physiological sensors, and behavioral logs) to create more sophisticated and reliable triggering mechanisms.
- **Investigate long-term effects**: Conduct longitudinal studies in authentic educational settings to examine how different scaffolding modes influence skill development, retention, and transfer over time.
- **Personalize EMME content**: Develop adaptive EMME systems that can generate or select video content based on the learner's current task context, error type, or demonstrated knowledge gaps.

## Conclusion

This study set out to examine whether scaffold initiation mode, who controls when EMME is delivered, determines its effectiveness in supporting novice programmers' performance. The answer is unambiguous: it does. All three EMME conditions outperformed the no-scaffold control, confirming that gaze-based scaffolding reliably enhances debugging performance when delivered in any form. However, human-mediated initiation, whether teacher- or learner-controlled, consistently produced higher performance than automated physiological triggering, establishing initiation mode as a critical and previously unexamined design variable for EMME and adaptive learning technologies more broadly. The automated triggering failure carries a specific and practically important message for system designers. Sustained low pupillary activity during complex problem-solving is an inherently ambiguous signal that cannot distinguish productive fluent processing from genuine impasse. Triggering support in response to low mental effort risks interrupting engagement that is already working, rather than supporting engagement that has broken down. Single-metric physiological triggering is not a sufficient basis for adaptive scaffold delivery in complex ill-structured domains, and future systems must combine richer behavioral, physiological, and task-state indicators to approximate the judgment that teachers and learners exercise naturally. Perhaps the most consequential finding for educational practice is that EMME eliminated the performance advantage of prior programming knowledge across all initiation modes. Learners with less prior knowledge benefited as much from EMME as their more knowledgeable peers, suggesting that gaze-based scaffolding can function as an equalizing resource, providing all learners with access to expert attentional and strategic processes that prior knowledge would otherwise confer. In introductory programming education, where expertise gaps persistently disadvantage novice learners, this democratizing potential may be EMME's most significant contribution. When appropriately timed and human-controlled, EMME does not merely support debugging; it levels the field on which debugging is learned.